# Sensing as a Service (S²aaS): Buying and Selling IoT Data

Charith Perera

The Internet of Things (IoT) [1] envisions the creation of an environment where everyday objects (e.g. microwaves, fridges, cars, coffee machines, etc.) are connected to the internet and make users' lives more convenient. It will also lead users to consume resources more efficiently.

Read More

Over the past few years, a large number of IoT solutions have come to the IoT marketplace [2]. Typically, each solution, consisting of one or more Internet Connected Objects (ICO), is designed to perform a single or minimal number of tasks (primary usage). For example, a smart sprinkler may only be activated if the soil moisture falls below a certain level in a garden. Further, smart plugs allow users to control electronic appliances (including legacy appliances) remotely or create automated schedules. Such automation not only brings convenience to users but also reduces resource wastage (e.g. through efficient planning and predictions).

The data collected by each of these solutions are used by them and stored in access controlled silos. After the primary usage, data are either thrown away or locked down in independent data silos. There is a significant amount of knowledge hidden in these silos that can be used to improve our lives (including behaviours, habits, and life patterns) and reduce wastage through efficient resource consumption. To discover such knowledge, it is essential to analyse data stuck in independent silos together on a large scale [3].

There are three main barriers to achieving this:
- Data owners do not have much control over their data and their data are locked in silos managed by products and services companies.
- Data owners only have access to their own data which has little value when it comes to knowledge discovery.
- Data owners do not know how to discover knowledge from raw data.

To overcome these barriers, we propose a Sensing as a Service (S²aaS) model [4]. It is a vision and a business model that promotes data exchange (i.e. trading) between data owners and data consumers.

Imagine a world where data owners (who own IoT solutions) get rewarded (e.g. money, loyalty points, gifts, vouchers, bitcoin, actionable advice [5], etc.) for sharing (i.e. trading) their personal data (collected by IoT products). From the other end, companies (i.e. data consumers) get to understand their customers (i.e. data owners) better. As a result, companies will be able to optimize their business operations by saving costs and create new products and services to fit individual customer needs. Data consumers may recover their data acquisition costs through business process optimization and increased customer (i.e. data owners) satisfaction. Data consumers can be government or not-for-profits as well [4].

The S$^2$aaS model provides stimuli for everyone to adopt IoT solutions and to participate in data trading as it helps data owners to recover the cost (at least in part) they invested on purchasing, deploying, and maintaining IoT solutions. However, the S$^2$aaS model is not going to be the primary reason that someone would want an IoT solution in their home. Instead, the convenience and the efficiency that IoT solutions bring would be the motivating factor. However, S$^2$aaS may act as a secondary motivation.

Today, we see a glimpse of data trading efforts. For example, Google Opinion Reward[1] and Survey.com[2] are mobile applications that selectively present survey questionnaires to users. Users get paid for answering the questionnaire surveys. It is important to note that users are getting paid for answering the surveys. Surveys like this have issues such as accuracy of answers, difficulty in asking lots of questions (users get bored quickly despite the fact that they are getting paid), difficulty in getting answers to data that users may not remember (e.g. how many times did the user drink coffee over the last month), and so on. Therefore, we can imagine how much value is hidden in the data captured by different IoT solutions. Another example is a start-up called Datacoup[3] that promises 8 USD per month[4] in return for trading personal data.

**A use case**

*Jane* is a restaurant manager who works in shifts. She lives alone in her own house. She has three different IoT products in her house. She has a context-aware thermostat that controls indoor temperature based on her preferences. She also has a smart coffee machine that automatically brews coffee when she gets up in the morning so by the time she arrives in the kitchen coffee is ready for her. Lastly, *Jane* has bought a smart activity monitor that monitors her exercise patterns, food intake, step counts, goals, and so on. These products are manufactured by three different companies and work independently.

*TastyCoffee* is a coffee products manufacture. They are keen to know how people like *Jane* consume coffee (e.g. patterns, amounts, etc.). *TastyCoffee* want to know whether there are any external factors that influence coffee consumption such as weather, temperature, workout patterns, etc. For example, *TastyCoffee* would like to discover any consumer patterns (e.g. *whether people tend to drink less coffee on a day with fewer workouts*). Currently, the only way that they could discover this kind of information is through user surveys and focus group studies. However, such methods are time consuming, inaccurate and expensive to carry out. However, if *TastyCoffee* can access *Jane's* silo (and thousands of other similar users) which consists of data recorded from all three of her IoT products (smart thermostat, smart coffee machine, activity monitor), they will be able to understand *Jane* (also thousands of other similar users) better and optimize their product supply chain. Such optimization will allow *TastyCoffee* to reduce costs and wastage, which would increase profits. Further, such data will help *TastyCoffee* to improve their product lines and introduce new products to the market rapidly, which will also lead to strengthening of their brand value. From *Jane's* perspective, rewards (e.g. voucher, discounts) she receives from *TastyCoffee* motivate her to participate and trade data in the S$^2$aaS model.

---

[1] https://play.google.com/store/apps/details?id=com.google.android.apps.paidtasks&hl=en
[2] https://play.google.com/store/apps/details?id=com.survey.android&hl=en
[3] https://datacoup.com/
[4] https://www.technologyreview.com/s/524621/sell-your-personal-data-for-8-a-month/

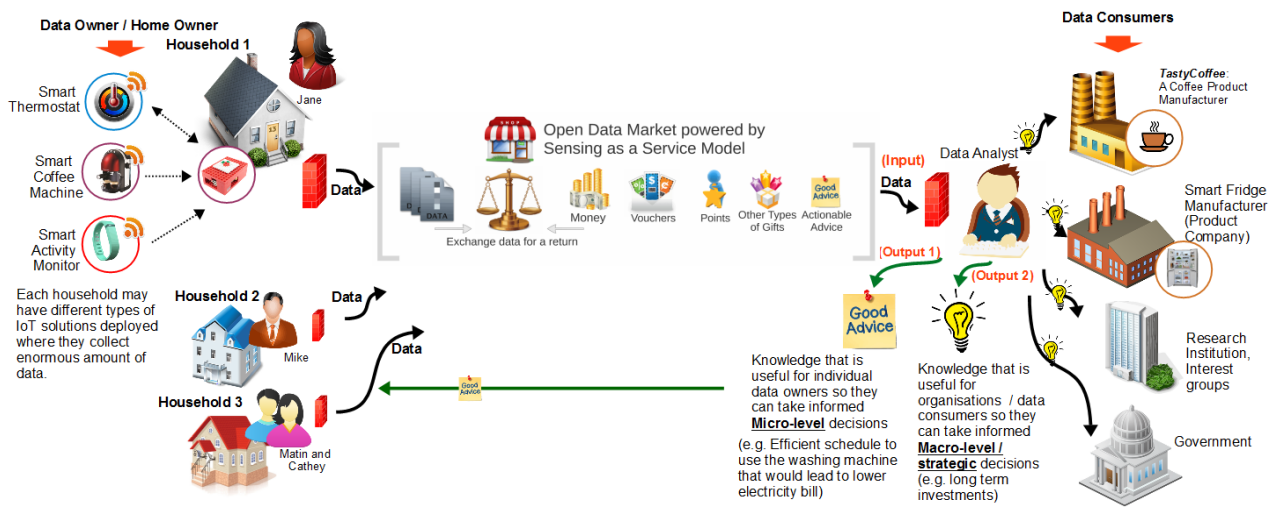

The figure summarises our discussion so far. The proposed S$^2$aaS model eliminates the previously mentioned barriers:

- Data owners will have full control of their data and they get to decide when and with whom they want to trade their data.
- Data consumers can acquire IoT data from a larger number of data owners through the data trading process, so they can use data analytics at scale to discover useful knowledge and insights.
- Data owners do not need to know how to analyse data or discover knowledge. Data consumers (i.e. third party services) will do it for them.

**Research challenges**

The main challenge is to develop the S$^2$aaS model in such a way that data owners can easily participate in data trading. Most of the data owners who engage in S$^2$aaS model are expected to be non-technical users. Therefore, all the user interactions and interfaces should be easy to understand and use. Attracting a large user base is vital for the success of the S$^2$aaS model. We conceptualize the challenges in building an S$^2$aaS model broadly under three themes: 1) S$^2$aaS Ecosystem and Infrastructure, 2) Setting up and Configuration, and 3) Data Trading and Negotiation.

One of the main challenges is to build the necessary *infrastructure and ecosystem* to support the S$^2$aaS model. It is envisioned that the S$^2$aaS model will be built on top of the IoT infrastructure (currently being built) for the most part. However, there are significant differences between the IoT and S$^2$aaS models. More specifically, the S$^2$aaS model requires additional infrastructure capabilities and novel interaction mechanisms. In the S$^2$aaS model, data consumers request data from data owners. The main challenges are how can data consumers make requests in such a way that non-technical data owners would be able to respond efficiently and effectively, and what constitutes a data request. Apart from direct data requests, how data consumers may advertise their interests to acquire certain types of data is also an important question to tackle.

Another challenge is *setting up and configuration* at the data owners' end. Ideally, when a data owner first brings an IoT solution home, there has to be a way to perform the initial configuration without requiring significant technical expertise. AllJoyn[5] can be considered as an initial framework that can be extended towards achieving this goal. The challenge is to find ways to minimise data owners' engagement in the initial configuration process. From the manufacturers' perspective, the question is how to build IoT solutions in such a way that they can easily be integrated into an $S^2$aaS model and open up the data, that is being collected and processed by each solution, for third parties at the request of data owners.

*Data trading and negotiation* are two main features that reside at the heart of the $S^2$aaS model. The challenge is to evaluate data requests made by data consumers and generate risk-reward analysis reports so the data owners can make informed data trading decisions. Visually representing risk-reward analyses in such a way that they are detailed enough for data owners to be informed accurately, but simple enough to be easily and quickly understood, is an important feature towards the success of the $S^2$aaS model. Another challenge is to decide what kind of controls should be given to data owners during both the negotiation and post-trading stages. The data buying and selling processes should be simple enough to take place repeatedly without requiring significant amounts of input and time from data owners. Finally, what aspects of a data trading transaction are negotiable and what is not, is also an important question to tackle. Baarslag et al. [6] has provided some insights towards data trading negotiations.

Even though some preliminary steps are being taken to address these challenges, the bulk of the research and development is yet to be done in order to fully realise the vision of the $S^2$aaS model. We hope the research challenges discussed here will help to develop the research agendas of the IoT community.

For more information, please download the ebook "Sensing as a Service for Internet of Things: A Roadmap" for free: https://leanpub.com/sensingasaservice

**Keywords**: Internet of Things, Sensing as a Service, Data Trading, Value Creation, Data Marketplace

---

[5] https://allseenalliance.org/framework

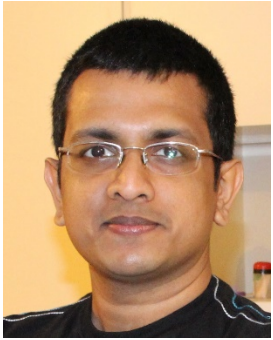

**Charith Perera** is a Research Associate at The Open University, UK. Currently, he is working on the Adaptive Security and Privacy (ASAP) research programme. He received his BSc (Hons) in Computer Science from Staffordshire University, UK and MBA in Business Administration from the University of Wales, Cardiff, UK and PhD in Computer Science at The Australian National University, Canberra, Australia. Previously, he worked at the Information Engineering Laboratory, ICT Centre, CSIRO. His research interests are Internet of Things, Sensing as a Service, Privacy, Middleware Platforms, and Sensing Infrastructure. He is a member of both IEEE and ACM. Contact him at [www.charithperera.net](www.charithperera.net) charith.perera@ieee.org